\documentstyle[12pt]{article}

\makeatletter
\def\@sect#1#2#3#4#5#6[#7]#8{\ifnum #2>\c@secnumdepth
    \def\@svsec{}\else
    \refstepcounter{#1}\edef\@svsec{\csname
    the#1\endcsname.\hskip 1em }\fi
    \@tempskipa #5\relax
    \ifdim \@tempskipa>\z@
    \begingroup #6\relax
    \@hangfrom{\hskip #3\relax\@svsec}{\interlinepenalty \@M #8\par}
    \endgroup
    \csname #1mark\endcsname{#7}\addcontentsline
    {toc}{#1}{\ifnum #2>\c@secnumdepth \else
     \protect\numberline{\csname the#1\endcsname}\fi
           #7}\else
    \def\@svsechd{#6\hskip #3\@svsec #8\csname #1mark\endcsname
          {#7}\addcontentsline
          {toc}{#1}{\ifnum #2>\c@secnumdepth \else
     \protect\numberline{\csname the#1\endcsname}\fi
           #7}}\fi
     \@xsect{#5}}
\def\label#1{\@bsphack\if@filesw {\let\thepage\relax
   \xdef\@gtempa{\write\@auxout{\string
   \newlabel{#1}{{\thesection.\@currentlabel}{\thepage}}}}}\@gtempa
   \if@nobreak \ifvmode\nobreak\fi\fi\fi\@esphack}
\def\@eqnnum{(\thesection.\theequation)}
\def\section{\setcounter{equation}{0} \@startsection {section}{1}{\z@}{-3.5ex
   plus -1ex minus -.2ex}{2.3ex plus .2ex}{\Large\bf}}
\newcount\@minsofar
\newcount\@min
\newcount\@cite@temp
\def\@citex[#1]#2{%
\if@filesw \immediate \write \@auxout {\string \citation {#2}}\fi
\@tempcntb\m@ne \let\@h@ld\relax \def\@citea{}%
\@min\m@ne%
\@cite{%
  \@for \@citeb:=#2\do {\@ifundefined {b@\@citeb}%
    {\@h@ld\@citea\@tempcntb\m@ne{\bf ?}%
    \@warning {Citation `\@citeb ' on page \thepage \space undefined}}%
{\@minsofar\z@ \@for \@scan@cites:=#2\do {%
  \@ifundefined{b@\@scan@cites}%
    {\@cite@temp\m@ne}
    {\@cite@temp\number\csname b@\@scan@cites \endcsname \relax}%
\ifnum\@cite@temp > \@min
    \ifnum\@minsofar = \z@
      \@minsofar\number\@cite@temp
      \edef\@scan@copy{\@scan@cites}\else
    \ifnum\@cite@temp < \@minsofar
      \@minsofar\number\@cite@temp
      \edef\@scan@copy{\@scan@cites}\fi\fi\fi}\@tempcnta\@min
  \ifnum\@minsofar > \z@ 
    \advance\@tempcnta\@ne
    \@min\@minsofar
    \ifnum\@tempcnta=\@minsofar 
      \ifx\@h@ld\relax
        \edef \@h@ld{\@citea\csname b@\@scan@copy\endcsname}%
    \else \edef\@h@ld{\ifmmode{-}\else--\fi\csname b@\@scan@copy\endcsname}%
      \fi
    \else \@h@ld\@citea\csname b@\@scan@copy\endcsname
          \let\@h@ld\relax
  \fi 
\fi}%
\def\@citea{,\penalty\@highpenalty\,}}\@h@ld}{#1}}
\def\appendixname{Appendix}
\def\appendix{\par
  \def\pre@section{\appendixname{} }
  \setcounter{section}{1}
  \@addtoreset{equation}{section}
  \def\thesection{\Alph{section}}
  \def\theequation{\arabic{equation}}}
\makeatother
\begin{document}
\def\t{\theta}
\def\T{\Theta}
\def\w{\omega}
\def\ov{\overline}
\def\a{\alpha}
\def\b{\beta}
\def\g{\gamma}
\def\s{\sigma}
\def\l{\lambda}
\def\wt{\widetilde}
\def\di{\displaystyle}
\def\sn{\mbox{sn}}
\def\cd{\mbox{cd}}
\def\cn{\mbox{cn}}
\def\dn{\mbox{dn}}
\def\tn{\mbox{tn}}

\begin{titlepage}
\vspace*{-4ex}
\null \hfill February 1995 \\
\vskip 3 true cm
\begin{center}
{\bf\Large Modified tetrahedron equation
and related 3D integrable models,II}\\[14ex]
{\large H. E. Boos}\footnote{ e-mail: boos@mx.ihep.su} \\ [6ex]
Institute for High Energy Physics,\\
Protvino, Moscow Region, Russia \\ [2ex]
\end{center}
\vskip 2.6cm
\begin{abstract}
This work is a continuation of  paper \cite{BMS}
where the Boltzmann weights for the N-state integrable spin
model on the cubic lattice has been obtained only numerically.
In this paper we present the analytical formulae for this
model in a particular case. Here the Boltzmann
weights depend on six free parameters including the elliptic
modulus. The obtained solution allows to construct a
two-parametric family of the commuting two-layer
transfer matrices. Presented model is expected to be  simpler
for a further investigation in comparison with a more general
model mentioned above.
\end{abstract}
\end{titlepage}

\addtolength{\unitlength}{-0.5\unitlength}
%


\section{Introduction}

In  paper \cite{BMS} a new family of N-state integrable three
dimensional two-layer models on cubic lattice has been  formulated.
The weight functions of this model have the composite structure. Namely,
they consist of eight more elementary weights which depend on eight indices
and four parameters $x$. These weights are a generalization used in
\cite{MS} of the weights for Baxter-
Bazhanov model \cite{BB1}-\cite{BB2}.  They have the
"Body-Centered-Cube" (BCC) structure
firstly introduced by Baxter in \cite{B2}  (more detailed \cite{B3})
 for Zamolodchikov's model with N=2 (\cite{Z1}-\cite{Z2}).
In  papers \cite{MS}-\cite{MSS} the idea to use a pair of
so-called modified
tetrahedron equations instead of a single tetrahedron one has
been proposed.  Paper \cite{BMS} is a further generalization of this
idea. Contrary to \cite{MS}-\cite{MSS} the elementary
weights should
satisfy only one of two sets of modified tetrahedron equations.
It     allows  to find a wider solution
which provides the integrability of the above mentioned  composite model.
It may be parameterized by the
elliptic functions which depend on the "angle-like"
parameters and one more parameter $k$ - the elliptic modulus.
In these terms each elementary weight depends on
three "angle-like" parameters and modulus $k$
instead of the four parameters $x$.
Any two weights corresponding to the neighbouring cubes in the lattice
have one coinciding "angle-like" parameter and opposite elliptic modulae:
$k$ and $-k$.
Rather cumbersome
analysis has shown us that the solution of the tetrahedron
equations for the composite         model has twelve free parameters
including elliptic modulus. This result has been obtained only numerically.
We have no  analytical formulae for this general case.

Our aim in this paper is to
consider
a simple particular case which could be more useful for a
calculation of the statistical sum and other physical quantities.
At the end of Section 6 of \cite{BMS}  an
example of such a model has been considered but it seems to be trivial.
In this paper we present another particular case which is probably
more substantial.

The paper
is organized as follows: In Section 2 we  remind the reader of the
basic formulae
and notations used in papers
\cite{KMS1}-\cite{BMS}.
 In Section 3 we give all necessary constraints and their solution for
 one composite weight $\cal W$. Also, we mention one interesting
property of this solution for our particular case:\
$Z$-invariance for the composite weights \cite{B4}.
This property works only in two-layer level  for the general case.
In Section 4 we discuss the
necessary
conditions for the existence of the intertwining composite weights
for $\cal W$ and $\cal W'$. The resulting formulae are collected in
the Appendix.

\section{Formulation of the model}

We would like to recall the basic notations used
in \cite{KMS1}-\cite{BMS}.
First of all let us remind the reader
that we shall consider a spin model on the
cubic lattice. In the Baxter-Bazhanov case the Boltzmann
weights corresponding
to the cells of this lattice are the same. This model may be called
homogeneous. Each Boltzmann weight depends on  eight
indices and parameters $x$ (see Fig.1):

\begin{picture}(600,265)
\put(0,50){
\begin{picture}(500,200)
\multiput(140,0)(120,0){2}{\line(0,1){120}}
\multiput(140,0)(0,120){2}{\line(1,0){120}}
\multiput(140,0)(0,120){2}{\line(-1,1){60}}
\put(80,180){\line(1,0){120}}\put(80,180){\line(0,-1){120}}
\put(200,180){\line(1,-1){60}}
\multiput(200,180)(0,-20){6}{\line(0,-1){12}}
\multiput(80,60)(20,0){6}{\line(1,0){12}}
\multiput(255,5)(-30,30){2}{\line(-1,1){20}}
\multiput(140,0)(120,0){2}{\circle*{10}}
\multiput(140,120)(120,0){2}{\circle*{10}}
\multiput(80,60)(120,0){2}{\circle*{10}}
\multiput(80,180)(120,0){2}{\circle*{10}}
\put(300,80){\large $=\quad W(a|e,f,g|b,c,d|h;\{x\})$}
\put(150,10){$e$}\put(270,10){$d$}\put(150,100){$a$}
\put(270,100){$f$}\put(212,186){$b$}\put(92,190){$g$}
\put(92,65){$c$}\put(212,65){$h$}
\end{picture}
}
\put(320,0){\bf Fig. 1}
\end{picture}

Up to some face factors we have
\begin{eqnarray}
&W(a|e,f,g|b,c,d|h; \{x\}) =&\nonumber \\
&\biggl\{{\di\sum_{\sigma\in Z_N}}{\di  w(x_3, x_{13},x_1|d,h+\sigma)
w(x_4,x_{24},x_2|a,g+\sigma)\over
\di w(x_6,x_{56},x_5|e,c+\sigma)w(x_8/\w,x_{78},x_7|f,b+\sigma)}
\biggr\}_0
&                                             \label{2}
\end{eqnarray}
where the lower index ``$0$'' after the curly brackets implies
that the expression in the curly brackets is divided by itself with all
exterior spin variables equated to zero.
The parameters $x_i$ and $x_{ij}$ satisfy the Fermat condition:
\begin{equation}
 x_{ij}^N = x_i^N-x_j^N.                                   \label{3}
\end{equation}
 In (\ref{2}) we used the following notations:
\begin{eqnarray}
w(x,y,z|k,l) &=& w(x,y,z|k-l) \Phi (l),\nonumber \\
w(x,y,z|k) &=& \prod_{s=1}^{k} \frac{y}{z-x \omega^s},\nonumber\\
\Phi (l) &=& \omega^{l(l+N)/2},\nonumber\\
\w &=& \exp(2\pi i/N)           \label{4}
\end{eqnarray}
and
\begin{equation}
x^N+y^N=z^N,                                                    \label{5}
\end{equation}
$l$ and $k$ are  elements of $Z_N$.

Below we shall use the following normalization conditions:
\begin{equation}
 x_3 = x_4 = x_6 = x_8 = 1.                  \label{6}
\end{equation}
In fact, each weight $W(a|e,f,g|b,c,d|h) $  depends only on four
continuous parameters $ x_1,x_2,x_5,x_7$ which in the case of the
Baxter-Bazhanov model are connected with each other
by the following constraint:
\begin{equation}
x_1\> x_2\> = x_5\> x_7.             \label{BBCON}
\end{equation}
The weights functions of such  form satisfy the tetrahedron equation
(\cite{KMS1}-\cite{KMS2}):

\begin{eqnarray}
\sum_{d}
&W(a_4|c_2,c_1,c_3|b_1,b_3,b_2|d)\> W'(c_1|b_2,a_3,b_1|c_4,d,c_6|b_4)&
\nonumber\\
\times&W''(b_1|d,c_4,c_3|a_2,b_3,b_4|c_5)\>
 W'''(d|b_2,b_4,b_3|c_5,c_2,c_6|a_1)&\nonumber\\
=\sum_{d}
&W'''(b_1|c_1,c_4,c_3|a_2,a_4,a_3|d)\>
W''(c_1|b_2,a_3,a_4|d,c_2,c_6|a_1)&
\nonumber\\
\times&W'(a_4|c_2,d,c_3|a_2,b_3,a_1|c_5)\>
W(d|a_1,a_3,a_2|c_4,c_5,c_6|b_4).&
                                                             \label{TE}
\end{eqnarray}
where  weights $W',W'',W'''$ depend on their own sets of
four parameters $\{x'_i\},\{x''_i\},\{x'''_i\}$ , accordingly.

We  write (\ref{TE}) more briefly in the shorthand notations:
\begin{equation}
W \>  W' \> W'' \>  W''' = W''' \>     W'' \> W' \>     W. \label{TE1}
\end{equation}

Below we shall not demand the condition (\ref{BBCON}).
So, each Boltzmann weight depends on  four independent
variables $x$.

 Authors of \cite{MS}-\cite{MSS} have proposed to consider a pair
of the tet\-ra\-hed\-ron-like
equations instead of a single one (\ref{TE1}):
\begin{equation}
W \> \ov W' \> W'' \> \ov W''' =
W''' \> \ov W'' \> W' \> \ov W \label{MTE1}
\end{equation}
and
\begin{equation}
\ov W \> W' \> \ov W'' \> W''' =
\ov W''' \>  W'' \> \ov W' \>  W \label{MTE2}
\end{equation}
where $\ov W$ depends on generally speaking some new variables $\ov x_i$.
 Equations of the form (\ref{MTE1}) or (\ref{MTE2}) are called
 modified tetrahedron equations.
This idea allows the commutativity of two-layer transfer matrices for the
cubic lattice with a "chess" structure. The composite Boltzmann weight
which corresponds to the $2\times 2\times 2$ cube of this lattice
satisfies the
tetrahedron equation.

In our previous
paper \cite{BMS} we have generalized this idea in a following
way. We have considered the composite weight $\cal W$
which consists of eight generally
speaking different
 elementary weights of the form (\ref{2}) as  shown
in  Fig.2:

\begin{picture}(600,400)
\put(200,65){
\begin{picture}(300,300)
\multiput(0,100)(0,110){3}{\line(5,-4){125}}
\multiput(220,320)(-110,0){2}{\line(5,-4){125}}
\multiput(0,100)(62.5,-50){2}{\line(0,1){220}}
\multiput(0,320)(62.5,-50){2}{\line(1,0){220}}
\multiput(125,0)(0,110){3}{\line(1,0){220}}
\multiput(125,0)(110,0){3}{\line(0,1){220}}
\multiput(5,100)(20,0){11}{\line(1,0){10}}
\multiput(5,210)(20,0){11}{\line(1,0){10}}
\multiput(67.5,160)(20,0){11}{\line(1,0){10}}
\multiput(67.5,50)(20,0){11}{\line(1,0){10}}
\multiput(220,315)(0,-20){11}{\line(0,-1){10}}
\multiput(282.5,265)(0,-20){11}{\line(0,-1){10}}
\multiput(110,315)(0,-20){11}{\line(0,-1){10}}
\multiput(172.5,265)(0,-20){11}{\line(0,-1){10}}
\multiput(282.5,265)(0,-20){11}{\line(0,-1){10}}
\multiput(220,210)(25,-20){5}{\line(5,-4){20}}
\multiput(110,210)(25,-20){5}{\line(5,-4){20}}
\multiput(220,100)(25,-20){5}{\line(5,-4){20}}
\multiput(110,100)(25,-20){5}{\line(5,-4){20}}
\multiput(0,100)(110,0){3}{\circle*{10}}
\multiput(0,210)(110,0){3}{\circle*{10}}
\multiput(0,320)(110,0){3}{\circle*{10}}
\multiput(62.5,50)(110,0){3}{\circle*{10}}
\multiput(62.5,160)(110,0){3}{\circle*{10}}
\multiput(62.5,270)(110,0){3}{\circle*{10}}
\multiput(125,0)(110,0){3}{\circle*{10}}
\multiput(125,110)(110,0){3}{\circle*{10}}
\multiput(125,220)(110,0){3}{\circle*{10}}
\multiput(319,40)(0,1){3}{\line(1,0){100}}
\multiput(319,40)(0,1){3}{\line(2,1){20}}
\multiput(319,40)(0,1){3}{\line(2,-1){20}}
\multiput(345,92)(0,1){3}{\line(1,0){40}}
\multiput(340,92)(-15,0){6}{\line(-1,0){10}}
\multiput(340,93)(-15,0){6}{\line(-1,0){10}}
\multiput(340,94)(-15,0){6}{\line(-1,0){10}}
\multiput(255,92)(0,1){3}{\line(2,1){20}}
\multiput(255,92)(0,1){3}{\line(2,-1){20}}
\multiput(319,160)(0,1){3}{\line(1,0){100}}
\multiput(319,160)(0,1){3}{\line(2,1){20}}
\multiput(319,160)(0,1){3}{\line(2,-1){20}}
\multiput(210,290)(0,1){3}{\line(1,0){100}}
\multiput(210,290)(0,1){3}{\line(2,1){20}}
\multiput(210,290)(0,1){3}{\line(2,-1){20}}
\multiput(30,270)(0,1){3}{\line(-1,0){100}}
\multiput(30,270)(0,1){3}{\line(-2,1){20}}
\multiput(30,270)(0,1){3}{\line(-2,-1){20}}
\multiput(30,130)(0,1){3}{\line(-1,0){100}}
\multiput(30,130)(0,1){3}{\line(-2,1){20}}
\multiput(30,130)(0,1){3}{\line(-2,-1){20}}
\multiput(90,180)(0,1){3}{\line(-1,0){120}}
\multiput(90,180)(0,1){3}{\line(-2,1){20}}
\multiput(90,180)(0,1){3}{\line(-2,-1){20}}
\multiput(90,70)(0,1){3}{\line(-1,0){120}}
\multiput(90,70)(0,1){3}{\line(-2,1){20}}
\multiput(90,70)(0,1){3}{\line(-2,-1){20}}
\put(430,150){\Large $W_f$}\put(430,30){\Large $W_d$}
\put(396,82){\Large $W_h$}\put(321,280){\Large $W_b$}
\put(-110,120){\Large $W_c$}\put(-110,260){\Large $W_g$}
\put(-70,170){\Large $W_a$}\put(-70,60){\Large $W_e$}
\end{picture}}
\put(200,0){{\bf Fig. 2} A composite weight ${\cal W}$.}
\end{picture}

\begin{eqnarray}
{\cal W}=\sum & W(\{x_a\}) W(\{x_b\}) W(\{x_c\}) W(\{x_d\}) &\nonumber \\
              & W(\{x_e\}) W(\{x_f\}) W(\{x_g\}) W(\{x_h\}) &,
                     \label{1}
\end{eqnarray}
where sum $\sum $ is implied to be over one internal index.

In order to satisfy the tetrahedron equation for such  composite weights
we have to consider sixteen modified tetrahedron equations
for  elementary weights of the form (ref.(6.3) of \cite{BMS}):
\begin{equation}
W \> W' \> W'' \> W'''
= \ov W'''\> \ov W'' \> \ov W' \> \ov W.   \label{OTE}
\end{equation}
We would like to note that there is no  such  pair of  equations
among them which would
be "conjugated" to each other as a pair of (\ref{MTE1})
and (\ref{MTE2}). If it was so we would be forced to restrict ourselves
to
the model proposed in \cite{MS}, \cite{MSS} .  Unfortunately, this system
of sixteen modified tetrahedron equations
                              appears to be too complicated to be solved
analytically. We have succeeded in finding the solution to this
system only numerically which depends on twelve free parameters. Also, a
particular case of this model has been considered at the end of Section 6 of
\cite{BMS}. In particular, the opposite elementary weights within the
composite one are connected with each other by the inversion
for that case.

Now we shall consider another particular case with the following
constraints
on the opposite elementary weights:

\begin{equation}
W_a = \ov W_h ,\  W_b = \ov W_e,
\ W_c = \ov W_f , \  W_d = \ov W_g.  \label{21}
\end{equation}

The same relations will be implied to be valid for another
composite weights $\cal W'$, $\cal W''$,$\cal W'''$.
Now the  system of the sixteen modified tetrahedron equations
mentioned above reduces to the following system of the eight ones:
\begin{eqnarray}
 W_g\>\ov W'_g\>W''_g\>\ov W'''_g &=&  W'''_g\> \ov W''_g\>
W'_g\>\ov W_g , \;\;
\ov W_g\> W'_e\>\ov W''_e\> W'''_e =  \ov W'''_e\>  W''_e\>
\ov W'_e\> W_g , \nonumber\\
 W_e\>\ov W'_e\>W''_f\>\ov W'''_f &=&  W'''_f\> \ov W''_f\>
W'_e\>\ov W_e , \;\;
 \ov W_e\> W'_g\>\ov W''_h\> W'''_h =  \ov W'''_h\>  W''_h\>
\ov W'_g\> W_e , \nonumber\\
 W_f\>\ov W'_h\>W''_h\>\ov W'''_e &=&  W'''_e\> \ov W''_h\>
W'_h\>\ov W_f , \;\;
 \ov W_f\> W'_f\>\ov W''_f\> W'''_g =  \ov W'''_g\>  W''_f\>
\ov W'_f\> W_f , \nonumber\\
 W_h\>\ov W'_f\>W''_e\>\ov W'''_h &=&  W'''_h\> \ov W''_e\>
W'_f\>\ov W_h , \;\;
 \ov W_h\> W'_h\>\ov W''_g\> W'''_f =  \ov W'''_f\>  W''_g\>
\ov W'_h\> W_h . \nonumber\\
&&     \label{ST}
\end{eqnarray}

So, our problem has been reduced to  finding the solution
to this system
for 32 elementary weights which depend on  their
own sets of
four parameters $x$ (see (\ref{2})). Below we shall
consider this problem in
 more details.

It is more convenient to solve the system of  modified equations
(\ref{ST})
in a number of steps. First of all let us note that four modified
equations among (\ref{ST}) have the form of (\ref{MTE1}) while four
residual ones are of the form (\ref{MTE2}). So, our first step is to
solve only one modified equation such as (\ref{MTE1}) or (\ref{MTE2}).
Then we are going        to     resolve
those constraints
                                                    which
connect with each other the parameters only of one composite
weight $\cal W$.
The same can be done for others $\cal W'$, $\cal W''$ ,$\cal W'''$.
The next step will be to find the necessary conditions on
the parameters
of the composite weights $\cal W$, $\cal W'$ which
provide the existence of
the intertwining weights $\cal W$ and $\cal W'$.  Our last
 step reduces
to applying the
solution obtained in the first step to the modified equations
(\ref{ST}) and to
finding the elementary components for $\cal W''$ and $\cal W'''$.

\section{Solution of the modified tetrahedron equation }
Let us consider the modified tetrahedron equation (\ref{MTE1}).
We need some additional notations for one elementary weight $W$ (\ref{2})
which depends on the four parameters $x_i$ as  was mentioned above.
Namely , let us introduce the following notations:
\begin{equation}
X_1 = x_1^N ,\quad X_2 = x_2^N,\quad  X_5 = x_5^N ,
\quad X_7 = x_7^N,       \label{7}
\end{equation}
\begin{equation}
X_{13}=x_{13}^N, \quad X_{24}=x_{24}^N, \quad X_{56}=x_{56}^N,
\quad  X_{78}=x_{78}^N,
\end{equation}
and taking into consideration  conditions (\ref{6})
\begin{equation}
X_{13}=X_1-1,
\quad  X_{24}=X_2-1,\quad  X_{56}=X_5-1, \quad  X_{78}=X_7-1.
\end{equation}
Then we can
introduce a set of the four parameters $\{m,J_i \ (i=1,2,3)\}$
instead of $X_i$:
\begin{equation}
m^2 =\frac{X_1 X_2}{X_5 X_7} , \quad
J_1 = \frac{X_7}{X_2}, \quad  J_2 = \frac{X_5}{X_2},
\quad J_3 = \frac{X_{56} X_{78}}{X_{13} X_{24}}.              \label{9}
\end{equation}
Note that $m$ is connected with the elliptic modulus $k$ introduced
in \cite{BMS} by the formula:
$$
m = \frac{1-k}{1+k}
$$
Also, it is convenient to introduce three more values
\begin{equation}
I_i = m^2 J_i.                                          \label{10}
\end{equation}
These notations are slightly different from those which were used in
\cite{BMS}.
These variables are different for the different weights.
For the validity of the modified tetrahedron equations these variables
should satisfy some definite relations which will be considered below.
In order to make it more descriptive it is convenient to
associate  variables $J_i$  with  three cube's
faces joined with a vertex "a" and $I_i$ with  three opposite faces
joined with  vertex "h" as  shown in Fig.3.

\begin{picture}(720,320)
\put(0,70){
\begin{picture}(500,200)
\multiput(140,0)(120,0){2}{\line(0,1){120}}
\multiput(140,0)(0,120){2}{\line(1,0){120}}
\multiput(140,0)(0,120){2}{\line(-1,1){60}}
\put(80,180){\line(1,0){120}}\put(80,180){\line(0,-1){120}}
\put(200,180){\line(1,-1){60}}
\multiput(140,0)(120,0){2}{\circle*{10}}
\multiput(140,120)(120,0){2}{\circle*{10}}
\put(80,60){\circle*{10}}
\multiput(80,180)(120,0){2}{\circle*{10}}
\put(150,10){$e$}\put(270,10){$d$}\put(150,100){$a$}
\put(270,100){$f$}\put(212,186){$b$}\put(92,190){$g$}
\put(92,65){$c$}
\put(50,115){\vector(1,0){60}}
\put(230,30){\vector(-1,1){30}}
\put(170,210){\vector(0,-1){60}}
\put(140,210){$J_1$}
\put(30,100){$J_2$}
\put(220,50){$J_3$}
\end{picture}
}
\put(370,70){
\begin{picture}(500,200)
\multiput(140,0)(120,0){2}{\line(0,1){120}}
\multiput(140,0)(0,120){2}{\line(1,0){120}}
\multiput(140,0)(0,120){2}{\line(-1,1){60}}
\put(80,180){\line(1,0){120}}\put(80,180){\line(0,-1){120}}
\put(200,180){\line(1,-1){60}}
\multiput(140,0)(120,0){2}{\circle*{10}}
\multiput(140,120)(120,0){2}{\circle*{10}}
\put(80,60){\circle*{10}}
\multiput(80,180)(120,0){2}{\circle*{10}}
\put(150,10){$b$}\put(270,10){$g$}\put(150,100){$h$}
\put(270,100){$c$}\put(212,186){$e$}\put(92,190){$d$}
\put(92,65){$f$}
\put(50,115){\vector(1,0){60}}
\put(230,30){\vector(-1,1){30}}
\put(170,210){\vector(0,-1){60}}
\put(140,210){$I_1$}
\put(30,100){$I_2$}
\put(220,50){$I_3$}
\end{picture}
}
\put(360,20){\bf Fig. 3}
\end{picture}

  The connection with the "angle-like" variables $ \theta_i$
introduced in \cite{BMS} is as follows:

\begin{equation}
     J_i = \frac{1}{m}\biggl( \frac{cn(\theta_i,k)-dn(\theta_i,k)}
{cn(\theta_i,k)+dn(\theta_i,k)}\biggr).                     \label{11}
\end{equation}
In \cite{BMS}, all algebraic relations which provide the
validity of the modified tetrahedron equation (\ref{MTE1})
where done.
One can choose from them those relations
 which contain the parameters of $W,W',W'',W'''$ only:
\begin{eqnarray}
J_1 &=& J'''_2, \quad J_2=J'_2, \quad I_3=1/J''_2,\nonumber\\
I'_1 &=& 1/J'''_3, \quad J'_3=J''_3, \quad J''_1=J'''_1    \label{16}
\end{eqnarray}
and
\begin{equation}
X_{24} \>  X''_{24} = X'_{24} \>  X'''_{24}.    \label{17}
\end{equation}
Below we
shall call all relations similar to (\ref{17}) as  relations of type
II. They connect together parameters of all weights.
We are interested only in a particular case:
\begin{equation}
m = m' = m'' = m'''.  \label{mmm}
\end{equation}
The analysis
of the equations (\ref{16}-\ref{mmm}) is a bit cumbersome,  but
rather straightforward and can be easily done by $\it MATHEMATICA $.
It can be shown that we have two different solutions.
 One of them seems to be meaningless and we shall discuss
 only the second one. Latter has six free parameters including the
modulus. We will not write down the manifest formulae.
We would like to note only that the variables
$X''_i$ and $X'''_i$ of the weights $W''$ and $W'''$, accordingly,
can be expressed rationally through the variables $X_i$ and $X'_i$
of $W$ and $W'$.
The residual relations considered in \cite{BMS} give us the variables for
$\ov W,\ov W',\ov W'',\ov W'''$.
Below we show them only for $W$, residual
variables can be obtained analogically:
\begin{eqnarray}
\ov m & = & 1/m , \quad \quad ( \ov k = -k ) \nonumber \\
\ov J_i & = & I_i, \quad \quad \ov I_i  =  J_i \nonumber \\
\ov X_1  &=& X_1 {\di (J_3 m^2 -1)\over \di m^2 (J_3-1)},\quad
\ov X_2  = X_2 {\di (J_3 m^2 -1)\over \di m^2 (J_3-1)},\nonumber\\
\ov X_5  &=& X_5 {\di (J_3 m^2 -1)\over \di (J_3-1)},\quad
\ov X_7  = X_7 {\di (J_3 m^2 -1)\over \di  (J_3-1)}.\label{14}
\end{eqnarray}
We shall call the substitution $W\rightarrow\ov W$ as $T$-transformation.
If two
weights $W$ and $W'$ correspond to the cubes which join each other
by  one face as it is shown in Fig.4, than we have
\begin{equation}
m'=\frac{1}{m}, \quad J'_2 = I_2, \quad I'_2 = J_2.
\end{equation}
The relations of this kind we will call as  relations of type I.
They connect
 variables only of the neighbouring weights in the lattice.

\begin{picture}(720,260)
\put(0,70){
\begin{picture}(500,200)
\multiput(140,0)(120,0){3}{\line(0,1){100}}
\multiput(140,0)(120,0){2}{\line(1,0){120}}
\put(140,0){\line(-1,1){60}}
\multiput(80,160)(120,0){2}{\line(1,0){120}}
\multiput(140,100)(120,0){2}{\line(1,0){120}}
\multiput(140,100)(120,0){3}{\line(-1,1){60}}
\put(80,160){\line(0,-1){100}}
\multiput(140,0)(120,0){3}{\circle*{10}}
\multiput(140,100)(120,0){3}{\circle*{10}}
\put(80,60){\circle*{10}}
\multiput(80,160)(120,0){3}{\circle*{10}}
\put(50,95){\vector(1,0){60}}
\put(30,80){$J_2$}
\put(200,60){$W$}
\put(320,60){$W'$}
\end{picture}
}
\put(300,20){\bf Fig. 4}
\end{picture}

The equations (\ref{16}) are  relations of type I.
They connect together  the variables of  two weights.
Above we have only described  how we can perform the first stage.
Let us
repeat that  obtaining the manifest formulae is straightforward.

\section{"Internal" constraints \
on the composite weight}

In order to
perform our following  step we need some additional information.
First of all, it is convenient for us to rewrite the equation (\ref{17})
in a different form.
Let us suppose that the equations (\ref{16}) have already been satisfied.
Then the equation  of type II (\ref{17})
is equivalent to the four relations of the "edge-like" parameters
introduced in \cite{BMS}:
\begin{eqnarray}
a_0+a'_2-a''_2+a'''_2 &=& 0\nonumber\\
a_1-a'_1+a''_1+a'''_0 &=& 0\nonumber\\
a_2+a'_0-a''_3+a'''_3 &=& 0\nonumber\\
a_3-a'_3+a''_0+a'''_1 &=& 0,                  \label{18}
\end{eqnarray}
where $a_i$ and $a_0$ are determined by the following formulae:
\begin{eqnarray}
\tan{a_i/2} &=& \sqrt{\frac{sn(\alpha_0,k) sn(\alpha_i,k)}
{sn(\alpha_j,k) sn(\alpha_k,k)}}, \nonumber \\
\tan{a_0/2} &=& k\>\sqrt{sn(\alpha_0,k) sn(\alpha_i,k)
sn(\alpha_j,k) sn(\alpha_k,k)},        \label{12}
\end{eqnarray}
and $\alpha_{\mu}$ are "excesses" :
\begin{equation}
\di\a_0 = {\di\t_1+\t_2+\t_3\over 2} - {\cal K},\quad
\a_r = \t_r-\a_0,\label{13}
\end{equation}
where $\cal K$
is a complete elliptic integral of the first kind for the
modulus $k$.

Below we also  need  the transformation properties of  weights (\ref{2})
obtained in \cite{BB2},\cite{KMS1},\cite{MSS}
 for the rotation $\rho$
on $\pi/2$ around the vertical axis and  $\tau$ -reflection. These two
transformations are generating elements for a whole group of the cube's
symmetry.
Namely,  weight (\ref{2}) is invariant upon these transformations up
to some face factors if  variables $X_i$ are changed
by $X^{\rho}_i$ for
$\rho$-rotation and $X^{\tau}_i$ for $\tau $-reflection. Also, we need
$X^{\lambda}_i$ for the rotation $\lambda$ on $2\pi/3$ around the a-h
direction of the cube (Fig.2):
\begin{eqnarray}
& \rho : \{a,e,f,g,b,c,d,h\}\rightarrow\{g,c,a,b,f,h,e,d\} &\nonumber\\
& &\nonumber\\
& X_1^{\rho} = \frac{X_{56}}{X_{13}},\quad  X_2^{\rho}=\frac{X_7}{X_2}
\frac{X_{24}}{X_{78}},
\quad  X_5^{\rho}=\frac{X_1}{X_5}\frac{X_{56}}{X_{13}}
,\quad  X_7^{\rho}=\frac{X_{24}}{X_{78}};& \nonumber \\
& & \nonumber\\
& \tau : \{a,e,f,g,b,c,d,h\}\rightarrow\{a,f,e,g,c,b,d,h\} & \nonumber\\
& & \nonumber\\
& X_1^{\tau} = X_1,\quad  X_2^{\tau}=X_2,\quad
X_5^{\tau}=X_7,\quad  X_7^{\tau}=X_5; & \nonumber\\
& & \nonumber\\
& \lambda : \{a,e,f,g,b,c,d,h\}
\rightarrow\{a,g,e,f,d,b,c,h\} & \nonumber\\
& & \nonumber\\
& X_1^{\lambda} = \frac{X_1}{X_5}\frac{X_{78}}{X_{24}}
\frac{(X_5-X_2)}{(X_1-X_7)},\quad
X_2^{\lambda} = \frac{X_{13}}{X_{56}}\frac{(X_5-X_2)}{(X_1-X_7)} ,
& \nonumber\\
& X_5^{\lambda} = \frac{X_7}{X_2}\frac{X_{13}}{X_{56}}
\frac{(X_5-X_2)}{(X_1-X_7)},\quad
X_7^{\lambda} = \frac{X_{78}}{X_{24}}\frac{(X_5-X_2)}{(X_1-X_7)}.&
\label{15}
\end{eqnarray}

Now let us consider the eight modified tetrahedron equations (\ref{ST}).
The analysis of the previous Section may be applied for each of them.
So, we can write all relations of type I such as (\ref{16}) and all
relations  of type II such as (\ref{18}). Combining together
these relations                        one can obtain those equations
which are  pure "internal" constraints on the variables of
 composite weights.
$$
m_e = m_f = m_g = m_h = m,
$$
\begin{eqnarray}
& J_{g1} = J_{f1} ,
\quad J_{g2} = J_{e2} , \quad  J_{g3} = J_{h3},& \nonumber\\
& J_{e1} = J_{h1} ,\quad J_{f2} = J_{h2} ,
\quad J_{e3} = J_{f3} &     \label{22}
\end{eqnarray}
and
\begin{eqnarray}
g_0 & +& e_0 + f_0 +   h_0 = 0\nonumber\\
g_1 & +& f_1 = e_1 +   h_1\nonumber\\
g_2 & +& e_2 = f_2 +   h_2\nonumber\\
g_3 & +& h_3 = e_3 +   f_3.          \label{23}
\end{eqnarray}
where $\{J_{ei},I_{ei},e_{\mu}\}$, $\{J_{fi},I_{fi},f_{\mu}\}$,
$\{J_{gi},I_{gi},g_{\mu}\}$ and $\{J_{hi},I_{hi},h_{\mu}\}$ are variables
$J_i,I_i$ defined by (\ref{9}-\ref{10}) and "edge-like" parameters
defined
by (\ref{12}) for weights $W_e$, $W_f$, $W_g$ and $W_h$ accordingly .
The analogical relations to (\ref{22}-\ref{23}) should be valid for
other composite weights ${\cal W',\cal W'',\cal W'''}$.
This situation is shown in Fig.5.
The last four relations of
type II may be replaced by four relations of  variables
$X_e,X_f,X_g,X_h$ for weights $W_e,W_f,W_g,W_h$ accordingly:
\begin{eqnarray}
\frac{X_{e13}}{\ov X_{e24}}\frac{X_{f24}}{\ov X_{f13}}
\frac{X_{g78}}{\ov X_{g56}}\frac{X_{h56}}{\ov X_{h78}} =
\frac{X_{e24}}{\ov X_{e13}}\frac{X_{f13}}{\ov X_{f24}}
\frac{X_{g56}}{\ov X_{g78}}\frac{X_{h78}}{\ov X_{h56}} & = & \nonumber\\
=\frac{X_{e56}}{\ov X_{e78}}\frac{X_{f78}}{\ov X_{f56}}
\frac{X_{g24}}{\ov X_{g13}}\frac{X_{h13}}{\ov X_{h24}} =
\frac{X_{e78}}{\ov X_{e56}}\frac{X_{f56}}{\ov X_{f78}}
\frac{X_{g13}}{\ov X_{g24}}\frac{X_{h24}}{\ov X_{h13}} & = &
\frac{X_{g1}}{m^2 \ov X_{g1}}\frac{X_{e1}}{\ov X_{e1}}.       \label{24}
\end{eqnarray}

\begin{picture}(720,260)
\put(50,50){
\begin{picture}(500,200)
\multiput(140,0)(120,0){2}{\line(0,1){120}}
\multiput(140,0)(0,120){2}{\line(1,0){120}}
\multiput(140,0)(0,120){2}{\line(-1,1){60}}
\put(80,180){\line(1,0){120}}\put(80,180){\line(0,-1){120}}
\put(200,180){\line(1,-1){60}}
\put(110,30){\line(0,1){120}}
\put(110,150){\line(1,0){120}}
\put(80,120){\line(1,-1){60}}
\put(140,180){\line(1,-1){60}}
\put(140,60){\line(1,0){120}}
\put(200,0){\line(0,1){120}}
\multiput(120,190)(60,0){2}{\vector(0,-1){30}}
\multiput(150,160)(60,0){2}{\vector(0,-1){30}}
\multiput(60,80)(0,60){2}{\vector(1,0){30}}
\multiput(90,50)(0,60){2}{\vector(1,0){30}}
\multiput(190,10)(60,0){2}{\vector(-1,1){30}}
\multiput(190,70)(60,0){2}{\vector(-1,1){30}}
\put(90,190){$J_{g1}$}\put(150,190){$I_{e1}$}
\put(125,160){$I_{e1}$}\put(182,160){$J_{g1}$}
\put(50,120){$J_{g2}$}\put(80,85){$I_{f2}$}
\put(45,60){$I_{f2}$}\put(80,25){$J_{g2}$}
\put(150,70){$I_{g3}$}\put(210,70){$J_{e3}$}
\put(150,10){$J_{e3}$}\put(210,10){$I_{g3}$}
\end{picture}
}
\put(150,20){\bf Fig. 5}
\end{picture}

The consequence of  equations (\ref{22},\ref{24}) are the following
relations which appear to be very useful:
\begin{eqnarray}
X_{f1}  =  p \  X_{g1}
\frac{X_{e1}}{\ov X_{e1}},&& \ov X_{f1}  =  p \  X_{g1},
\nonumber\\
X_{f2}  =  q \
\frac{X_{e1}}{\ov X_{e1}},&& \ov X_{f2}  =  q ,\nonumber\\
X_{f5}  =  p \  X_{g5} \frac{X_{e1}}{\ov X_{e1}}, &&
\ov X_{f5}  =  m^2 p X_{g5},\nonumber\\
X_{f7}  =  q \  \frac{X_{g1}}{m^2 X_{g5}} \frac{X_{e1}}{\ov X_{e1}}, &&
\ov X_{f7}  =  q \  \frac{X_{g1}}{X_{g5}}, \nonumber\\
X_{h1}  =  p \  X_{e1}
\frac{X_{g1}}{\ov X_{g1}},&& \ov X_{h1}  =  p \  X_{e1},
\nonumber\\
X_{h2}  =  q \  \frac{X_{e2}}{X_{g2}}\frac{X_{g1}}{\ov X_{g1}},&&
\ov X_{h2}  =  q \  \frac{X_{e2}}{X_{g2}},\nonumber\\
X_{h5}  =  p \  X_{e2} \frac{X_{g5}}{X_{g2}}\frac{X_{g1}}{\ov X_{g1}},&&
\ov X_{h5}  =  p \  m^2 X_{e2} \frac {X_{g5}}{X_{g2}},\nonumber\\
X_{h7} = \frac{q}{m^2}\frac{X_{e1}}{X_{g5}}\frac{X_{g1}}{\ov X_{g1}},&&
\ov X_{h7}  =  q \  \frac{X_{e1}}{X_{g5}},                  \label{25}
\end{eqnarray}
where p and q are some new parameters.
One can see that  substitution of these expressions into  equations
(\ref{22}) gives us a rational constraint on p and q. After one of
the relations (\ref{24})
has been satisfied we have two different solutions
for p and consequently q. For one of these solutions all three residual
relations (\ref{24}) are valid automatically and we have a six-parametric
solution. For another one, three residual relations (\ref{24}) give us
one additional quadratic constraint and we have
five parametric solution. We should note that this solution contains
 the Baxter-Bazhanov \cite{BB1}-\cite{BB2} model and
Zamolodchikov's one \cite{Z1}-\cite{Z2} for $N=2$ as a particular case.
Unfortunately, we have not
succeeded in achieving a final result in that case.
 Below we will consider only the first case which is much easier .
In this case  the following expressions for p and q
can be obtained:
\begin{equation}
p = \frac{(X_{e1}^{\rho} - X_{g5}^{\lambda})}
{(X_{e5}^{\rho} - X_{g2}^{\lambda})}
\frac{X_{g2}}{X_{e2} X_{g1} X_{g5}},\quad
q = \frac{(X_{e7}^{\rho} - X_{g1}^{\lambda})}{(X_{e2}^{\rho} - X_{g7}^
{\lambda})}
\frac{X_{g5}}{m^2 X_{e5} X_{g7}}.                      \label{26}
\end{equation}
where $X_{ei}^{\rho}$
and $X_{gi}^{\lambda} $ can be defined by  applying
 the formulae (\ref{15}) to  $W_e$ and $W_g$.
It is easy to obtain a
manifest formulae for variables $X_{fi}$,$\ov X_{fi}$
and $X_{hi}$,$\ov X_{hi}$ through the six
independent  variables (for example, $X_{g1},X_{g2},X_{g5},X_{e1},
X_{e2}$ and $ m$)
using (\ref{14}),(\ref{15}),(\ref{25}) and (\ref{26}).

To conclude this Section we would like to mention a remarkable fact
concerning our solution. Namely, let us consider two neighbouring weights
within the composite
weight $\cal W$ (see Fig.1),for example, $W_e$ and $W_d$
which is now  equal to $\ov W_g$ . We can find the intertwining weights
$W''$ and $W'''$ for this pair. In order to do so
we should substitute $X_{ei}$ and $X_{gi}$
into the equations (\ref{15}),(\ref{16}) instead of $X_i$ and $X'_i$
accordingly and resolve them with respect to  $X''_i$ and $X'''_i$,
as it was described in the previous Section.
We can follow the same procedure for  pairs $ W_a = \ov W_h$ and $W_f$ ,
$ W_c=\ov W_f$ and $W_h$ , $W_g$ and $W_b = \ov W_e$
and obtain the $Y''_i$ and $Y'''_i$ , $Z''_i$ and $Z'''_i$, $V''_i$ and
$V'''_i$ accordingly.  It is interesting to note
the following relations :
\begin{eqnarray}
Z''_i & = & \ov X''_i ,\quad Y'''_i = \ov X'''_i ,  \nonumber\\
Y''_i & = & \ov V''_i ,\quad Z'''_i = \ov V'''_i.        \label{zinv}
\end{eqnarray}
One can check that the similar situation takes place for all pairs of
the neighbouring elementary weights inside the composite weight.
This property is
nothing else but  $Z$-invariance  generalized on the three dimensional
case (\cite{B4}).
One should note that this property  works for a general
model (beyond the condition  (\ref{21})) only in the two-layer level and
breaks within the composite weight. We did not demand the validity of
$Z$-invariance
for the model (\ref{21}). That is why we were very surprised
to observe it.

\section{Existence of the intertwining weights}

Let us consider
the question about the existence of the intertwining composite
weights $\cal W'' $ and $\cal W''' $ for two original weights
$\cal W$ and $\cal W' $ which can be associated with the two cubes
joining each other by the face as  shown in Fig.6:

\begin{picture}(720,300)
\put(0,70){
\begin{picture}(500,200)
\multiput(140,0)(60,0){5}{\line(0,1){120}}
\multiput(140,0)(0,60){2}{\line(1,0){240}}
\multiput(140,0)(0,60){3}{\line(-1,1){60}}
\put(140,120){\line(1,0){240}}
\put(80,180){\line(1,0){240}}\put(80,180){\line(0,-1){120}}
\multiput(80,180)(60,0){5}{\line(1,-1){60}}
\put(110,30){\line(0,1){120}}
\put(110,150){\line(1,0){240}}
\multiput(60,80)(0,60){2}{\vector(1,0){30}}
\multiput(90,50)(0,60){2}{\vector(1,0){30}}
\put(50,120){$J_{g2}$}\put(80,85){$I_{f2}$}
\put(45,60){$I_{f2}$}\put(80,25){$J_{g2}$}
\end{picture}
}
\put(200,50){$\cal W$}
\put(320,50){$\cal W'$}
\put(260,20){\bf Fig. 6}
\end{picture}

As was mentioned above each of the weights $\cal W$ and $\cal W'$ has
six free parameters (including modulus). Besides, we have three relations
of type I:
\begin{equation}
m = m',\quad J_{g2} = J'_{g2},\quad J_{f2} = J'_{f2}.      \label{27}
\end{equation}
It seems to
be natural that there are no another constraints on the parameters
of $\cal W$ and $\cal W'$. If it was so we would have three parametric
family of the commuting two-layer transfer matrices.
But  accurate analysis has shown us that it is not the case.
There is one more constraint which is necessary for the existence
of $\cal W''$ and $\cal W'''$. So, the commuting family is only
two- parametric.

After the analysis of the previous Section has been performed
it is enough to resolve the following
set of  twelve equations of type I:
\begin{eqnarray}
 J_{g1}  =  J'''_{g2} &,&
 \quad J_{g2} = J'_{g2}, \quad J_{g3} = 1/I''_{g2},
\nonumber\\
 J'_{g1}  =  1/I'''_{g3} &,& \quad J'_{g3} = J''_{g3},
\quad J''_{g1} = J'''_{g1} , \nonumber\\
J_{e1}  =  J'''_{f2} &,&
\quad J_{e3} = 1/I''_{f2}, \quad J'_{e1}=1/I'''_{e3},
\nonumber\\
 J'_{e3}  =  J''_{e3} &,& \quad J_{f2} = J'_{f2},
\quad J''_{e1} = J'''_{e1}  \label{28}
\end{eqnarray}
and four relations of type II:
\begin{eqnarray}
 X_{g24} \> X''_{g24} & = & X'_{g24} \> X'''_{g24}, \nonumber\\
 X_{e24} \> X''_{f24} & = & X'_{e24} \> X'''_{f24}, \nonumber\\
 X_{f24} \> X''_{h24} & = & X'_{h24} \> X'''_{e24}, \nonumber\\
 X_{h24} \> X''_{e24} & = & X'_{f24} \> X'''_{h24}.  \label{29}
\end{eqnarray}
Below we describe our final result.
Our choice of  independent variables is as follows: \\
modulus $m$, five parameters from $\cal W$: $X_{g1},X_{g2},
X_{g5},X_{e1},X_{e2}$
and two parameters from $\cal W'$: $X'_{g1},X'_{g2}$.
In order not to encumber the text we have collected the resulting
formulae for $X'_{ei}$ in the Appendix.

 Variables $\ov X'_{ei}$ can be obtained
from the formulae (\ref{14}) and definitions
(\ref{9}) applied to  weight $W'_e$.  Expressions for  $X'_{fi},
\ov X'_{fi}$ and $X'_{hi},\ov X'_{hi}$ can be extracted by
 substituting
already known values into the formulae (\ref{25}) for the
composite weight $\cal W'$.
Also, it is necessary to use the following expressions for $p'$ and $q'$:
\begin{equation}
p' = \frac{p}{S_1},\quad q'=\frac{q}{S_1}\>
\frac{X'_{g2}}{X_{g2}}. \label{33}
\end{equation}
Now we know all variables for  weights $\cal W$ and $\cal W'$ and
use the analysis of  Section 3 for  finding
all intertwining elementary weights from (\ref{ST}) using a known
solution of the equations (\ref{16}-\ref{17}).
It is remarkable to note that all "internal" constraints described in
Section 4 are satisfied automatically.
Let us
remind the reader
that we have solved all necessary equations for $N$'th powers
of the original variables $x$. Now we need to choose the right powers of
$\omega $
when extracting the $N$'th roots. We have done it for a general
situation and a particular case considered in Section 6 of \cite{BMS}.
This experience tells us that it can be done. Moreover, as a rule there is
some arbitrariness in this procedure.

\section{Conclusion}

In this paper
we have considered a particular case of the general two-layer
integrable model proposed in \cite{BMS}. In fact, we have discussed only
one out of two possible solutions.                   This solution    has
five free parameters and one modulus which is the same for all composite
weights. It is interesting to note that for this solution the
so-called $Z$-invariance
has been restored inside each composite weight while
for the general
model this property takes place only on the two-layer level.
 Finding of the intertwiners for a pair of the composite weights $\cal
W$ and $\cal W'$ appeared to be unexpectedly difficult. In spite of our
expectation the
commuting family of the transfer matrices has only two free
parameters (not three). Unfortunately, we have not clarified completely
the meaning of
an  additional constraint. The model discussed above is simple
enough and at the same time it has enough  free parameters.
So, we hope that this
model may be useful for study of it's thermodynamic properties.

\noindent
{\bf Acknowledgement}

\noindent
Author thanks Professor W. Zimmermann for his hospitality
at the Max-Plank-Institut f\"ur Physik in M\"unchen where
author was enabled to discuss this paper.
This work is supported by the International Scientific
Fund (INTAS), Grant No. RMM000. Author would like to thank
Yu.G. Stroganov for reading the paper
and making many useful remarks.
Also, author would like to thank V.V.Mangazeev ,   S.M.Sergeev
E. Seiler, M. Niedermaier, O. Ogievetsky, R. Flume and
V. Rittenberg for very fruitful discussions.
 Computer calculations were carried out using {\it Mathematica}.

\section{Appendix}

Here we present the resulting formulae for $X'_{ei}$. These formulae being
written in terms of the independent variables are rather cumbersome.
To simplify them  we need  more notations:
\begin{eqnarray}
y_g & = & X_{g5}\> X_{g13}\> X_{g24}\> (1 - I_{g3}),\nonumber\\
z_g & = & m^2 \> X_{g5}\> X_{g13}\> X_{g24}\> (1 - J_{g3}),\nonumber\\
y_e & = & X_{g2} \> X_{g5}\> X_{e13}\> X_{e24}\> (1 - I_{e3}),\nonumber\\
z_e & = & m^2 \> X_{g2}\> X_{g5}\>
X_{e13}\> X_{e24}\> (1 - J_{e3}),\nonumber\\
y'_g & = & X_{g2}\>X_{g5}\>
X'_{g13}\> X'_{g24}\> (1 - I'_{g3}),\nonumber\\
z'_g & = & m^2\>X_{g2}\>X_{g5}\>
X'_{g13}\> X'_{g24}\> (1 - J'_{g3}).\label{34}
\end{eqnarray}

Note that
\begin{equation}
\frac{\ov X_{g1}}{X_{g1}} =
\frac{y_g}{z_g} \quad \frac{\ov X_{e1}}{X_{e1}} =
\frac{y_e}{z_e} \quad \frac{\ov X'_{g1}}{X'_{g1}} = \frac{y'_g}{z'_g}.
\label{35}
\end{equation}
Also, let us introduce :
\begin{eqnarray}
\alpha & = & m^2\>X_{g5}\> y_g - X_{e1}\> z_g \nonumber\\
\beta & = & X_{g5}\> z_e - X_{e1}\> y_e \label{36}
\end{eqnarray}
and
\begin{eqnarray}
w_1 & = & X'_{g1}\>z_g - m^2\> X_{g5}\> y_g,\nonumber\\
w_2 & = & X_{e1}\>X_{g1}\>X_{g2} - m^2\> X_{e2}\>X_{g5}^2,\nonumber\\
w_3 & = & m^2\>X_{g5}\>y'_g - X_{g1}\>z'_g,\nonumber\\
w_4 & = & X_{e1}\>X_{g2}\>y'_g - X_{g5}\>X_{e2}\>z'_g,\nonumber\\
w_5 & = & X_{e24}\>(X_{g2} - m^2\>X_{g5})\>(X_{g2}-X_{e2}\>X_{g5})\>w_2 +
\nonumber\\
 & + & (X'_{g2} - X_{e2})\>X_{g5}\>
(X_{g1}\>X_{g2}\>z_e - m^2\>X_{e2}\>X_{g5}\>y_e).
\label{37}
\end{eqnarray}

After that our result looks as follows:
\begin{eqnarray}
X'_{e1} & = & S_1\>S_2 ,\quad X'_{e2} = X_{g2}\> S_1\> S_3, \nonumber\\
X'_{e5} & = &
\frac{X_{g5}}{X_{g2}}\>X'_{e2}, \quad X'_{e7} = \frac{X_{g2}}
{m^2 X_{g5}}\> X'_{e1},            \label{30}
\end{eqnarray}
where
\begin{equation}
S_1 = \frac{l_1}{r_1},
\quad S_2 = \frac{l_2}{r_2}, \quad S_3 = \frac{l_3}{r_3}
\label{31}
\end{equation}
and
\begin{eqnarray}
l_1 & = & X'_{g1}\ (X'_{g2}-X_{g2})\ y_e\ \alpha - X'_{g2}\ w_1\ \beta,
\nonumber\\
r_1 & = & X_{g5}\ (X'_{g2}-X_{g2})\ z_e\ \alpha - X_{g2}\ w_1\ \beta,
\nonumber\\
l_2 & = & w_1\ w_2\ y'_g - X_{e2}\ X_{g5}\ w_3\ \alpha,\nonumber\\
r_2 & = & X_{g2}\ (X_{e1}-X'_{g1})\ w_3\ y_g + w_1\ w_4,\nonumber\\
l_3 & = & X_{e2}\
X_{g2}\ w_3\ \beta - (X'_{g2} - X_{g2})\ w_2\ z_e\ y'_g,
\nonumber\\
r_3 & = & (m^2-1)\ X_{g2}\ X_{g5}\
(X_{e1}-X'_{g1})\ (X'_{g2}-X_{g2})\ w_5+
\nonumber\\
& + & X_{g2}\ (X'_{g2}-X_{e2})\ r_1 + l_3 .                 \label{32}
\end{eqnarray}


\begin{thebibliography}{**}

\bibitem{Z1}
A.B. Zamolodchikov, Zh. Eksp. Teor. Fiz. {\bf79} (1980) 641-664
[English trans.: JETP {\bf 52} (1980) 325-336].

\bibitem{Z2}
A.B. Zamolodchikov, Commun. Math. Phys. {\bf79} (1981) 489-505.

\bibitem{B1}
R.J. Baxter, Commun. Math. Phys. {\bf88} (1983) 185-205.

\bibitem{B2}
R.J. Baxter, Phys. Rev. Lett., Vol. 53,{\bf19} (1984) 1795-1798.

\bibitem{B3}
R. J. Baxter, Physica D, {8} (1986) 321 -- 347.

\bibitem{B4}
R.J. Baxter, Phys. Trans. Roy. Soc. (London),Vol.289\ A, {\bf1359} (1978)
315 - 346.

\bibitem{BB1}
V.V. Bazhanov, R.J. Baxter, J. Stat. Phys. {\bf 69} (1992) 453-485.

\bibitem{BB2}
V.V. Bazhanov, R.J. Baxter, J. Stat. Phys. {\bf71} (1993) 839.

\bibitem{KMS1}
R.M. Kashaev, V.V. Mangazeev,
Yu.G. Stroganov, Int. J. Mod. Phys. {\bf A8}
(1993) 1399-1409.

\bibitem{KMS2}
R.M. Kashaev, V.V. Mangazeev,
Yu.G. Stroganov, Int. J. Mod. Phys. {\bf A8}
(1993) 587-601.


\bibitem{MS} V.V. Mangazeev,
Yu.G. Stroganov: {\it Elliptic solution for
modified tetrahedron equations.}
Preprint IHEP 93-80, 1993, (HEP-TH/9305145),
to appear in Mod. Phys. Lett. A.


\bibitem{MSS}
V.V. Mangazeev, S.M. Sergeev, Yu.G. Stroganov,\\
 to appear in Int. J. Mod. Phys.

\bibitem{BMS}
H.E. Boos, V.V. Mangazeev, S.M. Sergeev, \\
  {\it Modified tetrahedron
   equation and related 3D integrable models.} \\
   Preprint IHEP 94-76, HEP-TH/9407146, to appear in  Int. J. Mod. Phys..

\end{thebibliography}
\end{document}